\documentclass[aip,graphicx,amsmath,amssymb,reprint]{revtex4-1}

\usepackage[utf8]{inputenc}
\usepackage[T1]{fontenc}
\usepackage{dcolumn}
\usepackage{graphicx}
\usepackage{textcomp}
\usepackage{mathptmx}

\draft % marks overfull lines with a black rule on the right

\begin{document}

%\preprint{}

\title[Tuning the superconducting transition of 
SrTiO$_3$-based 2DEGs with light]{Tuning the superconducting transition
of SrTiO$_3$-based 2DEGs with light}



\author{D. Arnold}
\affiliation{Institute for Solid-State Physics, Karlsruhe Institute of Technology, Karlsruhe, Germany}
\author{D. Fuchs}
\affiliation{Institute for Solid-State Physics, Karlsruhe Institute of Technology, Karlsruhe, Germany}
\author{K. Wolff}
\affiliation{Institute for Solid-State Physics, Karlsruhe Institute of Technology, Karlsruhe, Germany}
\author{R. Sch\"afer}
\affiliation{Institute for Solid-State Physics, Karlsruhe Institute of Technology, Karlsruhe, Germany}
\email{roland.schaefer@kit.edu}

%\thanks{}


\date{\today}

\begin{abstract}
	The resistivity of the two dimensional electron gas that forms at the
	interface of strontium titanate with various oxides is sensitive to
	irradiation with visible light.  In this letter we present data on the
	interface between the band gap insulators LaAlO$_3$ (LAO) and SrTiO$_3$
	(STO).  We operate a light emitting diode at temperatures below 1\,K
	and utilize it to irradiate the LAO/STO interface at ultra low
	temperatures.  On irradiation the resistance of this system is lowered
	continuously by a factor of five and the resistance change is
	persistent at low temperatures as long as the sample is kept in the
	dark.  This makes a characterization of transport properties in
	different resistive states over extended time periods possible.  Our
	pristine sample gets superconducting below 265\,mK.  The transition
	temperature $T_c$ shifts downwards on the persistent photo-induced
	lowering of the resistance.  The persistent photoconductance can be
	completely reverted by heating the structure above 10\,K in which case
	$T_c$ as well takes on its original value.  Thus very similar to field
	effect control of electron densities irradiation at low temperatures
	offers a versatile tuning knob for the superconducting state of
	STO-based interfaces which in addition has the advantage to be
	nonvolatile.
\end{abstract}

%\pacs{}% insert suggested PACS numbers in braces on next line

\maketitle %\maketitle must follow title, authors, abstract and \pacs

A two dimensional electron gas (2DEG) develops at the interface between
strontium titanate (STO) and a variety of different oxides\cite{Ohtomo2004,
Seo2007, Perna2010, Moetakef2011, Li2011, Lee2012, He2012, Annadi2012b,
Li2013, Chen2013, Xu2014} which offers a huge playground for many kind of
solid state phenomena (e.g.\ Ref. \onlinecite{Zubko2011}).  Here we focus on
low temperature transport properties which are known to be sensitive to
exposure of the 2DEG to visible light at room temperature.\cite{Huijben2006,
Lei2014, Li2015, Tebano2012}  Careful experiments keep therefore all samples
for a considerable time in a dark environment (typically 24 h) prior to cool
down to avoid photo-induced effects which otherwise lead to relaxation effects
and a drift of transport coefficients on amazingly long time
scales.\cite{Huijben2006} The effect of photo induced conductivity has of
course been addressed as an independent subject of interest.\cite{Yang2017,
Cheng2017, Yazdi-Rizi2017} It has been studied in recent years from room
temperature down to $T=1.5\,$K and was found to be persistent at low
temperature\cite{Cheng2017} in accordance with the aforementioned precautions
taken by many researchers.  The persistent charge carriers have been linked to
oxygen vacancies trapped at domain boundaries which develop below the so-called
antiferrodistortive transition at 105\,K.\cite{Yazdi-Rizi2017}

A salient feature of low temperature transport of STO-based 2DEGs is
superconducitvity found below $T=300\,$mK.\cite{Reyren2007, Biscaras2010,
Fuchs2014} The two dimensional confinement of the mobile carriers make it easy
to gate tune its density and in turn the conductance by the field
effect.\cite{Thiel2006}  For the superconducting transition temperature $T_c$ a
dome shaped structure in the phase diagram\cite{Caviglia2008, Gariglio2016} was
found, which resembles celebrated findings in the cuprates and bulk STO.
However, despite a decade of intense research the microscopic origin of the
shift in $T_c$ with gate tuning remains controversial, as the influence and
interplay of important parameters like disorder, inhomogeneity and spin-orbit
coupling is not fully understood.  It is therefore desirable to find new
control parameters altering transport characteristics.\cite{Fuchs2015}

In this letter we establish persistent photo-conductance as another tuning
knob for the superconducting transition temperature which might give further
insight in the nature of the superconducting state of the STO-based 2DEG.  We
operate a light emitting diode (LED) at dilution fridge temperatures.  By
stabilizing the temperature at $T=500\,$mK we can monitor the resistance
during irradiation by the LED and find a continuous reduction.  When the LED
is switched off the resistance is constant.  On the time scale of our
experiment (which in some cases extended for a period of more than a week) we
could not detect any change of resistance as long as the temperature stays
well below $T<1\,$K and the LED is switched off.  However, the resistance
change can be completely reverted by a controlled elevation of temperature to
$1\,$K $<T<15\,$K.  The phenomenon seams to be a rather general feature and
has so far been observed in LaAlO$_3$/SrTiO$_3$ (LAO/STO) heterostructures as
well as in the $\gamma-$Al$_2$O$_3$/SrTiO$_3$ system.  Here we present
exemplary data on a LAO/STO sample where we could reduce the resistance by a
factor of five.  The rate of resistance change depends on the radiant flux of
the LED and is to first order proportional to the LED current.  Adjusting the
radiant energy we can set the resistance value within the total tuning range
on purpose.  This statement is true for both directions of resistance change.
Resistance can be tuned downwards by light and upwards again by elevation of
temperature. 

The sample has been prepared by standard pulsed laser ablation using
TiO$_2$-terminated (001) SrTiO$_3$ substrates and LaAlO$_3$ single crystal
targets. Film deposition was done at an oxygen partial pressure of
$p(\text{O}_2) = 10^{-5}\,$mbar onto a substrate heated to $T_{\text{sub}}
=700\,^\circ$C. More details on sample preparation and patterning are described
elsewhere.\cite{Fuchs2017,Wolff2018} Electrical connections to the six arms of
a Hall bar geometry are made by ultrasound wire bonding with aluminum leads.

\begin{figure}
	\includegraphics[width=\columnwidth]{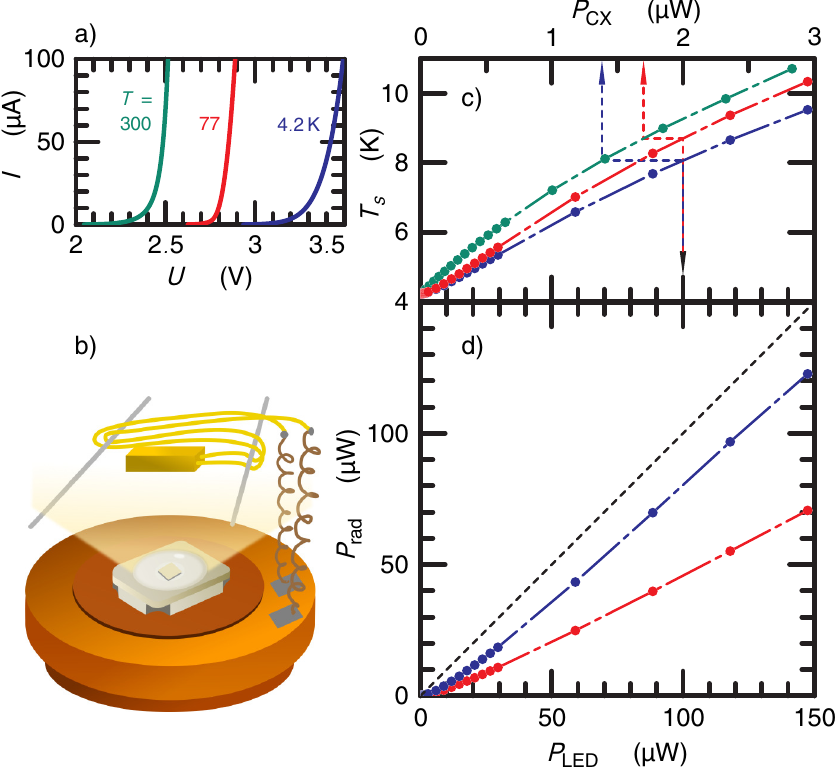}
	\caption{\label{fig:one} (a) I/U characteristics of the light emitting
	diode (LED) at different temperatures. (b) Schematics of the LED
	calibration experiment.  The radiant flux of the LED irradiates a
	thermometer chip (gold) supported by thin nylon fibres (gray).  The
	leads for measuring the thermometer resistance consist of thin
	manganin wires. (c) Temperature of the thermometer chip as a function
	of electrical power dissipated by the LED (lower scale labeled
	$P_{\text{LED}}$, red and blue data).  In the same diagram the
	temperature elevation due to self heating of the thermometer (upper
	scale labeled $P_{\text{CX}}$, green data) is shown.  With the help of
	this diagram $P_{\text{LED}}$ can be mapped onto the power dissipated
	by thermometer due to light absorption.  The difference between the
	red and the blue data is explained in the text. (d) Radiant flux of
	the LED as a function of effective input power at $T=4.2\,$K.  The
	data shown in blue represent the total flux, while data shown in red
	correspond to the part of the spectrum absorbed by a gold plated
	surface.}
\end{figure}

For the purpose of irradiating the Hall bar structure at $T<1\,$K we use a
white LED (OSRAM ''\textit{Golden DRAGON Plus}'', type LW
W5AM)\cite{osram:lw-w5am} with a radiation spectrum made up of a narrow
primary emission line centered around a wavelength of 460\,nm and a broad
photo luminescence (PL) band with a maximal intensity around 565\,nm.
The diode is intended to be used in the temperature range $-40\,^\circ°$C
$<T<125\,^\circ$C.\cite{osram:lw-w5am} By recording I/U characteristics in an
extended temperature range (Fig.\ \ref{fig:one} (a)) we gained confidence that
the LED works properly even below $T<1\,$K.  As expected, the forward voltage
shifts to higher voltages on cool down but this shift is rather moderate.  At
$T= 77\,$K we could operate the LED in an open dewar and check by visual
inspection that the emitted light did not show a noticeable shift in color.
However, to avoid excess heating when used at $T<1\,$K the LED has to be
operated at extremely low currents (below
$I_{\text{LED}}<I_{\text{max}}=50$\,\textmu{}A), owing to the limited cooling
power of our dilution fridge. $I_{\text{max}}$ is by more than three orders of
magnitude smaller than the minimal recommended diode current
100\,mA.\cite{osram:lw-w5am}  It is not clear how efficient electrical power
is converted into electromagnetic radiation in this situation. 

To gain some knowledge of the efficiency we performed the experiment sketched
in Fig.\ \ref{fig:one} (b) in which the LED is held at $T=4.2\,$K and a
resistive thermometer is mounted at a distance of 10\,mm with sufficient
thermal resistance to show a well resolved temperature rise when irradiated at
low diode currents.  The setup is installed on a general purpose puck of a
physical property measurement system (PPMS, Quantum Design) which provides the
temperature reservoir.  The LED is thermally anchored thoroughly to the puck
while the calibrated Cernox thermometer ($R_{\text{CX}}$) (Lake Shore
Cryotronics, CX-1050-SD-1.4L) is supported by two thin Nylon fibers
(80\,\textmu{}m diameter, actually dominating the thermal resistance) and
electrically connected by two 30\,\textmu{}m thick Manganin wires of about
150\,mm length.  When irradiated by the LED or self-heated by a sufficiently
high measurement current $I_{\text{CX}}$ the thermometer reaches an elevated
steady state temperature $T_s$ after a relaxation time of typically 30\,min.
We measure $T_s$ as a function of $I_{\text{CX}}$ and the diode current
$I_{\text{LED}}$.  A rise of $T_s$ as response to a diode current is clear
evidence for the presence of an irradiance flux absorbed by $R_{\text{CX}}$.  We
take the smallest forward voltage $U_{\text{LED}}=2.95\,$V at which we could
observe a temperature rise in $R_{\text{CX}}$ (corresponding to a diode current
$I_{\text{LED}}=100\,$nA) as a measure of the photon energy of the primary
emission line at $T=4.2\,$K.  The radiant energy of the LED is then strictly
bounded by the effective power $P_{\text{LED}}=I_{\text{LED}}\cdot2.95\,$V.  In
Fig.\ \ref{fig:one} (c) we display $T_s$ as function of $P_{\text{LED}}$ for
two different runs.  The blue data are recorded while the thermometer chip had
its original gold plated color.  It than absorbs predominantly the short
wavelength photons from the main emission line while the longer wavelength PL
photons are reflected.  In a first run (red) the thermometer chip was covered
by a black paint produced by mixing varnish (GE 7031) with carbon black.  In
this case the absorption includes the long wavelength part of the spectrum
resulting in an increased $T_s$ at identical $P_{\text{LED}}$ levels.  We also
recorded $T_s$ as a function of the dissipated power $P_{\text{CX}}$ due to
self heating when $I_{\text{CX}}$ is increased giving identical results in both
runs (green).  As indicated by the arrow headed lines in Fig.\ \ref{fig:one}
(c), we can relate $P_{\text{LED}}$ to $P_{\text{CX}}$ which gives a measure
for the irradiance flux absorbed by $R_{\text{CX}}$.  Finally, a geometry
factor $f$ can be deduced relating radiant power of the LED to irradiance flux
received by the surface of $R_{\text{CX}}$.  Actually, uncertainties in the
latter quantity dominates the systematic error of this experiment and limits
the accuracy of the final result presented in Fig.\ \ref{fig:one} (d) to about
20\%.  In this figure $f$ is taken into account to calculate the radiant flux
$P_{\text{rad}}$ which is plotted as a function of $P_{\text{LED}}$.  The black
dashed line represents the theoretical limit, while the blue and red data
represent the fraction of radiant flux dissipated by a black and gold plated
absorber, respectively.  Fig.\ \ref{fig:one} (d) reports an amazingly high
efficiency of the LED at 4.2\,K. The total radiant intensity of our light
source is about 500\,nW$/($\textmu{}A\,sr$)\cdot{}I_{\text{LED}}$ in forward
direction. The persistent photoconductance discussed in the rest of this letter
is most likely caused by the short wave length part of the spectrum (this is
the radiation absorbed by the gold plated $R_{\text{CX}}$). Its radiant
intensity is found to be 300\,nW$/($\textmu{}A\,sr$)\cdot{}I_{\text{LED}}$.

\begin{figure}
	\includegraphics[width=\columnwidth]{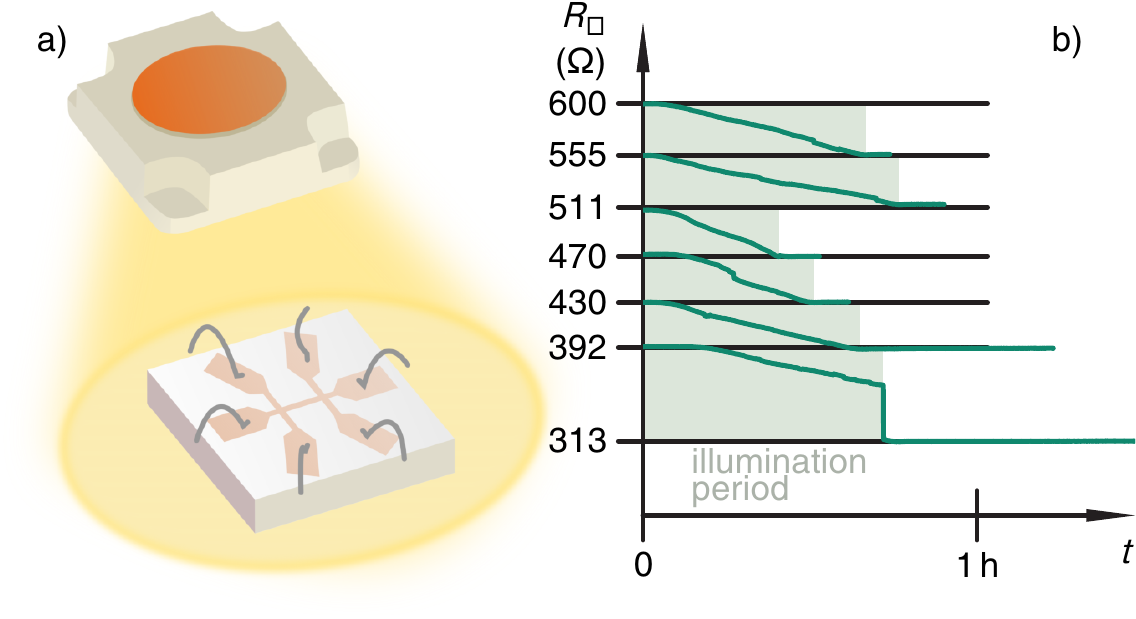}
	\caption{\label{fig:two} (a) Schematics of the LAO/STO sample
	structured in Hall bar geometry and irradiated by a LED.  The setup is
	operated in a dilution refrigerator at $T\le500\,$mK. (b) Sheet
	resistance $R_\Box$ vs.\ time at $T=500\,$mK.  During light exposure
	(period marked in green) $R_\Box$ drops continuously while it stays
	constant when the LED is switched off.  In some instances $R_\Box$
	display a sudden jump during illumination.  One such instance can be
	seen between $392\,\Omega$ and $313\,\Omega$.}
\end{figure}
After characterizing the light source, we describe in the rest of this letter
our main experiment sketched in Fig.~\ref{fig:two}~(a) which is installed in
a commercial dilution refrigerator (Oxford Instruments, MX250). The Hall bar
structured sample is thermally anchored at a sample stage. The temperature of
the sample stage $T_{\text{ss}}$ is PID regulated using a resistor chip $R_h$
as heater and can be controlled in the range
$12\,\text{mK}<T_{\text{ss}}<2\,$K. Reaching higher temperatures is difficult
without removing the $^3$He/$^4$He insert from the liquid helium bath.
Nevertheless, we managed to heat our sample to about
$T_{\text{ss}}\approx10\,$K by running our insert in a special operation mode
(see below). In this case the temperature is only weakly controlled.  The sheet
resistance $R_\Box$ of the sample is measured with a lock-in technique: An ac
current of amplitude $I_{\text{AC}}=9\,$nA which is reduced to
$I_{\text{AC}}=2\,$nA for temperature dependent $R_\Box(T)$ measurements is
sourced to the central strip of the Hall bar structure.  The resulting voltage
drop $U_{\text{AC}}$ over two side terminals is amplified by a home-build
amplifier and measured by the lock-in amplifier (Signal Recovery, Model 7265
DSP).  The length between the voltage terminals equals 12 times the width of
the central strip and $R_\Box=U_{\text{AC}}/(12I_{\text{AC}})$.  We carefully
checked that we stay in the linear regime of the current voltage
characteristics which made it necessary to reduce the measurement current close
the superconducting transition.  Hall resistance is recorded in fields up to
$8\,$T by measuring the voltage across two terminals opposite to each other and
supplies us with information on the sheet carrier concentration.

The LED light source is mounted at $\sim10\,$mm distance from the sample.  LED
currents of $100\,$nA $<I_{\text{LED}}<50\,$\textmu{}A are supplied by a
source/measure unit (Keithley Model 2400) via superconducting leads. Operating
the LED at higher power leads to a heat burden of the sample stage. In
principle $I_{\text{LED}}$ could replace the current sourced to $R_h$ in the
PID regulation circuit.  For the experiments described here we use $R_h$ for
temperature control and always stabilize the stage at $T_{\text{ss}}=500\,$mK
before slowly turning on $I_{\text{LED}}$. The PID regulation reacts to
increasing $P_{\text{LED}}$ by reducing the current sourced to $R_h$ keeping
$T_{\text{ss}}$ almost constant ($|\Delta{}T_{\text{ss}}|<5\,$mK). At
$I_{\text{LED}}=50\,$\textmu{}A the current through $R_h$ is set to zero by the
PID circuitry and heating is solely due to the dissipation of the LED.

\begin{figure*}
	\includegraphics[width=\textwidth]{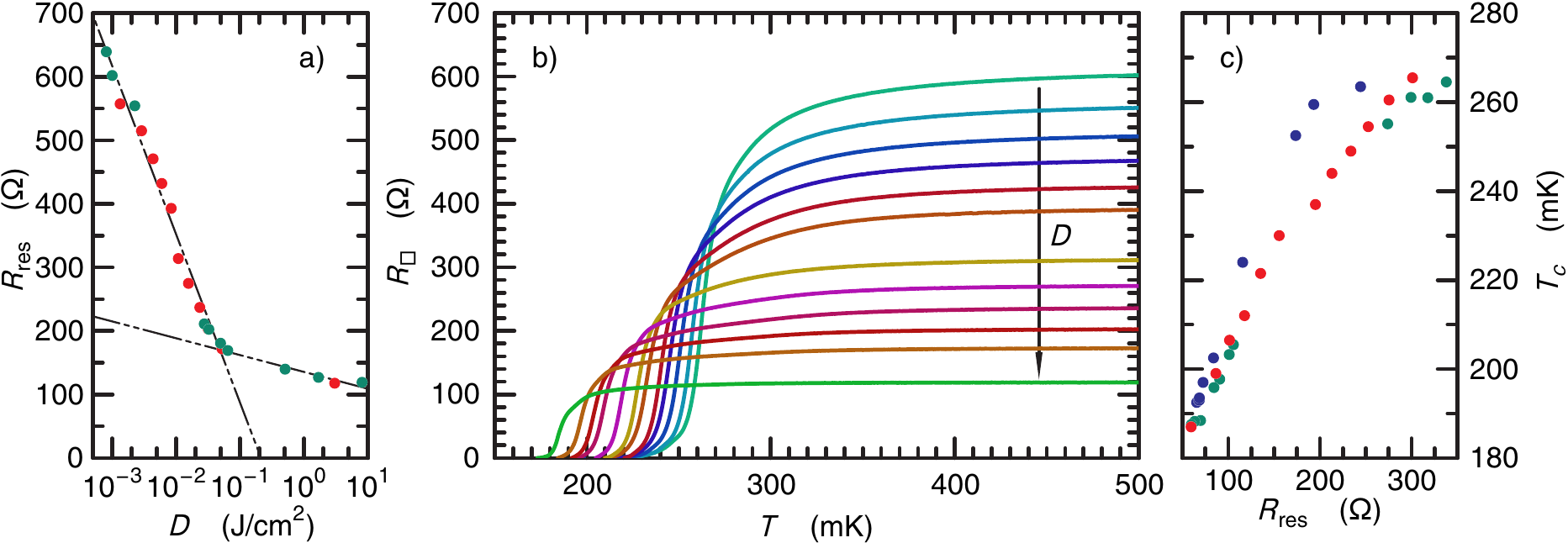}
	\caption{\label{fig:three} (a) Residual sheet resistance
	$R_{\text{res}}=R_\Box(500\,\text{mK})$ as function of radiant exposure
	$D$ in the short wavelength part of the LED spectrum. Red dots
	correspond to the irradiance sequence shown in (b). Green dots
	represent the result of a second sequence recorded after a temperature
	induced reset of $R_{\text{res}}$. Black lines are guide to the eyes
	explained further in the text.
	(b) Sheet resistance $R_\Box$ as a function of temperature $T$ for
	various residual resistances $R_{\text{res}}$.  The shown curves are
	measured successively while $R_{\text{res}}$ has been reduced in
	between by a controlled irradiation. The LED is off during the
	measurements.
	(c) Superconducting transition temperature $T_c$ as function of
	$R_{\text{res}}$. Red and green symbols as in (a), while blue dots are
	results during temperature controlled resistance increase.}
\end{figure*}

Constant $T_{\text{ss}}$ does not automatically guarantee a constant
temperature of the 2DEG at the LAO/STO interface of our sample.  Because of
the positive temperature coefficient of $R_\Box$ at $T=500\,$mK, an increase of
temperature of the 2DEG as response to irradiation would result in an increase
of $R_\Box$.  In the contrary, light exposure at constant $T_{\text{ss}}$
leads to a decrease of $R_\Box$. It alters the resistive state of our sample
and this alteration cannot be attributed to heating. Examples are given in in
Fig.\ \ref{fig:two} (b).  Within the green marked period the sample was
illuminated by setting $I_{\text{LED}}$ to values between $1$ and
$5\,$\textmu{}A.  During illumination $R_\Box$ decreases as a function of
time. As soon as we turn off $I_{\text{LED}}$, $R_\Box$ stabilizes at its
reduced, momentary value. As long as the LED remains switched off and
$T_{\text{ss}}\lesssim1\,$K, $R_\Box$ is a function of magnetic field and
temperature only and does not change over time. Initially,
$R_{\text{res}}\equiv{}R_\Box(T=500\,\text{mK})$ was found to be
$R_{\text{res}}=600\,\Omega$. Subsequently we altered the resistive state of
our sample in steps of about $42\,\Omega$. In Fig.~\ref{fig:three}~(a)
where $R_{\text{res}}$ is displayed as a function of the radiant exposure $D$,
this first series is represented by red symbols. The abscissa in this figure
is deduced from $\int I_{\text{LED}}\mathrm{d}t$ with the help of the LED
calibration presented earlier and corresponds to the total radiant energy per
unit area irradiated to the sample by the LED in the short wavelength part of
the spectrum. Most likely only high energetic photons are responsible for the
persistent conductance effect.\cite{Yazdi-Rizi2017}  However, we do not have
further evidence for this statement and additional experiments with
monochromatic LEDs are required for a proof. The total radiant exposure over
all wavelength is 70\% larger.

Light exposure leads to persistent reduction of $R_\Box$. However, the
resistance change can be reverted by heating the sample to
$T_{\text{ss}}>1\,$K. The rate of resistance \emph{increase}
depends in this case on temperature and speeds up considerable above
$T>4.2\,$K.  Unfortunately it is almost impossible to regulate temperatures in
the regime $T>2\,$K in our dilution fridge.  While we could reach temperatures
of the order $12\,$K $>T>10\,$K by thermally isolating\footnote{This is done by
pumping out the $^3$He/$^4$He-mixture.  An absorption pump at the condenser is
still kept at $T= 1.7\,$K to maintain good isolation from the $^4$He bath.
This, actually, limits the maximal sample temperature we can reach by heating
the mixing chamber.} and heating the mixing chamber this temperature is only
weakly controlled.  Nevertheless, we could convince ourselves, that
$R_{\text{res}}$ measured after cooling the sample down again after
heat treatment depends strongly on the peak temperature reached during the
procedure and only weakly on the time period it lasts.  The minimal resistance
$R_{\text{res}}=120\,\Omega$ we could achieve by irradiation is in this
respect related to the temperature $T=500\,$mK at which it is measured and
lower resistive states might in principle be reached by irradiation at lower
temperatures but relax back to $R_{\text{res}}=120\,\Omega$ if the sample is
heated to $T=500\,$mK. 

By controlled heating we are able to increase $R_{\text{res}}$ in small steps.
Again, at low temperature and while the LED is switched off, $R_\Box$ is a function of
magnetic field and temperature only and is absolutely stable over time. 
By heating the sample to $T_{\text{ss}}\approx12$\,K we could reach
resistance states with even slightly higher $R_{\text{res}}=675\,\Omega$ than
the initial value of $600\,\Omega$. Subsequently we lowered the resistance
again by utilizing the LED. This second run is shown as green symbols in
Fig.~\ref{fig:three}~(a) giving consistent results. Analyzing the
dependence of $R_{\text{res}}(D)$ gives the empirical results shown as black
lines in Fig.~\ref{fig:three}~(a). The experimental data are
well described by 
$$
R_{\text{res}}(D)=R_{\text{ini}}-\Delta{}R\cdot{}L_D,\quad
L_D\equiv10\log_{10}\left(\frac{D}{1(\text{mJ\,cm}^{-2})}\right)\,\text{dB},
$$
with different initial resistance values ($R_{\text{ini}}=617\,\Omega$, $215\,\Omega$) and
slopes ($\Delta{}R=26.5\,\Omega/$dB, $2.65\,\Omega/$dB) at lower and higher dose, respectively.

In Fig.\ \ref{fig:three} (b) we present $R_\Box(T)$ in the different
persistent resistance states of the first light induced reduction series
(corresponding to the red symbols in Fig.~\ref{fig:three}~(a)).  During
this measurements $I_{\text{LED}}=0$.  The curves in Fig.\ \ref{fig:three}
(b) display superconductivity at low temperatures with a transition temperature
$T_c$ which shifts downwards with decreasing $R_{\text{res}}$.  This effect is
summarized in Fig.~\ref{fig:three}~(c) where $T_c$ was estimated as the
temperature where $R_\Box(T_c)=R_{\text{res}}/2$.  In the initial sequence of
resistance reduction by illumination $T_c$ decreased monotonically from
$T_c=265\,$mK down to $T_c=187\,$mK.  The blue symbols represent the findings
when $R_{\text{res}}$ recovers due to controlled heating.
$R_{\text{res}}=490\,\Omega$ could be reached easily by heating the sample to
$T\approx11\,$K.  In this state $T_c$ fully recovers to $T_c=264\,$mK.  We then
kept the sample for an extended period of several days at $T\lesssim12\,$K and
could increase $R_{\text{res}}$ even further ($R_{\text{res}}=675\,\Omega$).
However, this had only a minor effect on $T_c$.  The final illumination
sequence with falling $R_{\text{res}}$ is shown as green symbols and shows the
reproducibility of our finding.

The observed reduction of $T_c$ with decreasing $R_{\text{res}}$ is typical for
the so called over-doped regime and indeed taking the resistance range of our
experiment ($120\,\Omega<R_\Box<600\,\Omega$) into account this finding is in
accord with published data\cite{Caviglia2008,Gariglio2016} on field effect
tuned resistivities.  Our Hall resistivity measurements show a zero field slope
($R_H=43\,\Omega/$T) which is almost independent of $R_{\text{res}}$ in
accordance with data presented in Ref.~\onlinecite{Joshua2012}.  Within the
limited field range of our magnet ($\pm8\,$T) we only see slight nonlinearities
at larger fields which get more pronounced as $R_{\text{res}}$ is lowered. This trend
has already been reported previously for LAO/STO in the overdoped regime. 

For now, the microscopic mechanism of light induced changes is absolutely
unclear.  Findings by other authors\cite{Yazdi-Rizi2017} differ in subtle
details.  We completely revert the photo-induced transport changes at
comperatively low temperatures ($T\approx12\,$K) while in Ref.\
\onlinecite{Yazdi-Rizi2017} a crossing of the antiferro-distortive transition
of STO at $T\approx105\,$K seems to be necessary to revert persistence in
photo-conductance.  The difference might route in a smaller photon energy in
our case.  The LED radiates at $h\nu\lesssim2.95\,$eV, which is of the order
but considerable smaller than the band gap of STO ($E_{\text{gap}}=3.2\,$eV).
Other experiments on persistent photo-conductance use UV light above the gap
energy. 

In summary, we present a setup to tune the transport behavior of STO-based
interfaces at low temperatures with light. The radiant intensity of a LED was
calibrated at $T=4.2\,$K as a function of current and utilized as the light source
below 1\,K.  Adjusting the radiant energy we were able to tune the residual
sheet resistance $R_\text{{res}}$ at 500 mK, while simultaneously changing the
superconducting transition temperature $T_c$.  We reported a monotonous
behavior of $T_c$ vs $R_\text{{res}}$ for different resitive states.  To
reverse the altered state we used heat treatment up to 12\,K.  Using visible
light at low temperatures we are introducing a new nonvolatile tuning
parameter on the superconductivity of the STO-based interfaces.

This paper was supported by the Deutsche Forschungsgemeinschaft (Grant SCHA
658/2-1 and FU 457/2-1).  We thank K.~Grube for fruitful discussion. 

%\nocite{*}
\bibliography{DArnold-RSchaefer}

\end{document}